\documentclass[reprint, amsmath,amssymb, aps, superscriptaddress, floatfix]{revtex4-1}

\usepackage[utf8]{inputenc}
\usepackage{graphicx}
\usepackage{dcolumn}
\usepackage{bm}
\usepackage{braket}
\usepackage{color}
\usepackage{units}
\usepackage{dsfont}
\usepackage{float}
\usepackage{hyperref}
\usepackage{xcolor}

\usepackage{tabularx}
\usepackage{mathrsfs}

\begin{document}
\title{Multiphoton interference in a single-spatial-mode quantum walk}

\author{Kate L. Fenwick}
\affiliation{Department of Physics, University of Ottawa, Advanced Research Complex, 25 Templeton Street, Ottawa ON Canada, K1N 6N5}
\affiliation{National Research Council of Canada, 100 Sussex Drive, Ottawa, Ontario K1A 0R6, Canada}
\author{Jonathan Baker}
\affiliation{Department of Physics, University of Ottawa, Advanced Research Complex, 25 Templeton Street, Ottawa ON Canada, K1N 6N5}
\author{Guillaume S. Thekkadath}
\affiliation{National Research Council of Canada, 100 Sussex Drive, Ottawa, Ontario K1A 0R6, Canada}
\author{Aaron Z. Goldberg}
\affiliation{National Research Council of Canada, 100 Sussex Drive, Ottawa, Ontario K1A 0R6, Canada}
\author{Khabat Heshami}
\affiliation{Department of Physics, University of Ottawa, Advanced Research Complex, 25 Templeton Street, Ottawa ON Canada, K1N 6N5}
\affiliation{National Research Council of Canada, 100 Sussex Drive, Ottawa, Ontario K1A 0R6, Canada}
\author{Philip J. Bustard}
\affiliation{National Research Council of Canada, 100 Sussex Drive, Ottawa, Ontario K1A 0R6, Canada}
\author{Duncan England}
\affiliation{National Research Council of Canada, 100 Sussex Drive, Ottawa, Ontario K1A 0R6, Canada}
\author{Fr\'ed\'eric Bouchard}
\email{frederic.bouchard@nrc-cnrc.gc.ca}
\affiliation{National Research Council of Canada, 100 Sussex Drive, Ottawa, Ontario K1A 0R6, Canada}
\author{Benjamin Sussman}
\affiliation{Department of Physics, University of Ottawa, Advanced Research Complex, 25 Templeton Street, Ottawa ON Canada, K1N 6N5}
\affiliation{National Research Council of Canada, 100 Sussex Drive, Ottawa, Ontario K1A 0R6, Canada}

\begin{abstract}
Multiphoton interference is crucial to many photonic quantum technologies. In particular, interference forms the basis of optical quantum information processing platforms and can lead to significant computational advantages. It is therefore interesting to study the interference arising from various states of light in large interferometric networks. Here, we implement a quantum walk in a highly stable, low-loss, multiport interferometer with up to 24 ultrafast time bins. This time-bin interferometer comprises a sequence of birefringent crystals which produce pulses  separated by 4.3\,ps, all along a single optical axis. Ultrafast Kerr gating in an optical fiber is employed to time-demultiplex the output from the quantum walk. We measure one-, two-, and three-photon interference arising from various input state combinations, including a heralded single-photon state, a thermal state, and an attenuated coherent state at one or more input ports. Our results demonstrate that ultrafast time bins are a promising platform to observe large-scale multiphoton interference.
\end{abstract}

\maketitle

\section*{Introduction}
Wave interference is ubiquitous in the physical world.  While often associated with signal degradation, interference is also of considerable utility. For example, controlled interference of acoustic waves is critical for instrument tuning~\cite{helmholtz2009sensations} and optimized performance in concert halls~\cite{dammerud2010stage}, while metrology frequently relies on interferometry and can reach exquisite levels of sensitivity, \textit{e.g.,} in measuring gravitational waves~\cite{abbott2016observation}. By contrast, the ``waves’’ of quantum mechanics interfere mathematically in Hilbert space. Quantum interference arises from the fact that quantum systems can remain in a superposition of many possible states, only until measurement. The unusual scaling of quantum interference differentiates it from classical interference, making it is a key ingredient for quantum computational speedup~\cite{stahlke2014quantum} and the distribution of quantum walks, as demonstrated here.

Quantum interference can be exploited in many platforms, including trapped atoms~\cite{graham2022multi} or ions~\cite{cirac2000scalable, kielpinski2002architecture}, nuclear magnetic resonance~\cite{jones2000geometric}, and superconducting qubits~\cite{gambetta2017building}; however, photonic platforms are naturally suited to this task, offering advantages such as the ability to operate at room temperature and exceptional resilience to decoherence~\cite{arute2019quantum, zhong2020quantum, madsen2022quantum}. Amplitude-splitting interferometers~\cite{michelson1887relative} remain the tool nonpareil for the generation and manipulation of optical interference. They rely on the action of a beam splitter, and are also well-suited to measure quantum interference: when two indistinguishable single photons are each incident at one of the two input ports of a beam splitter, the hallmark ``Hong-Ou-Mandel'' interference dip can be observed in coincident detection at the output~\cite{hong1987measurement,bouchard2020two}. This effect is the foundation of logic gates in linear optical quantum computing (LOQC)~\cite{knill2001scheme}. Large scale implementations of LOQC will require the propagation of multiphoton input states in reconfigurable interferometric networks~\cite{carosini2024programmable}. In this context, experimental investigations of quantum interference in large interferometric networks become increasingly relevant and valuable.

The quantum walk (QW)~\cite{aharonov1993quantum,kempe2003quantum}, the quantum mechanical analog of the random walk, provides an excellent platform for generating, manipulating, and observing quantum interference. Here, we use a photonic QW to manipulate and observe multiphoton interference for a variety of input states. Photonic QWs have been demonstrated experimentally, with interference in both the path and time degrees of freedom, in many different platforms, including bulk optics~\cite{do2005experimental,bouwmeester1999optical,broome2010discrete,kitagawa2012observation,cardano2015quantum,cardano2016statistical,cardano2017detection,nejadsattari2019experimental,geraldi2019experimental,d2020two,esposito2022quantum,di2023ultra,wang2024efficient,di2024manifestation}, fiber loops~\cite{schreiber2010photons,schreiber20122d,jeong2013experimental,lorz2019photonic,geraldi2021transient,bagrets2021probing,held2022driven,pegoraro2023dynamic}, fiber cavities~\cite{boutari2016large}, integrated photonics~\cite{peruzzo2010quantum,sansoni2012two,harris2017quantum,pitsios2017photonic,meany2015laser,grafe2016integrated}, and measurement-based approaches~\cite{de2024realization}. Nevertheless, certain limitations must be taken into account when selecting a platform to study multiphoton interference in a QW. The overall phase stability of the multi-mode interferometer must be high, the loss per step will necessarily be low, and each individual layer of the interferometric network should be readily programmable. 

\begin{figure*}[t!]
	\centering
		\includegraphics[width=1\textwidth]{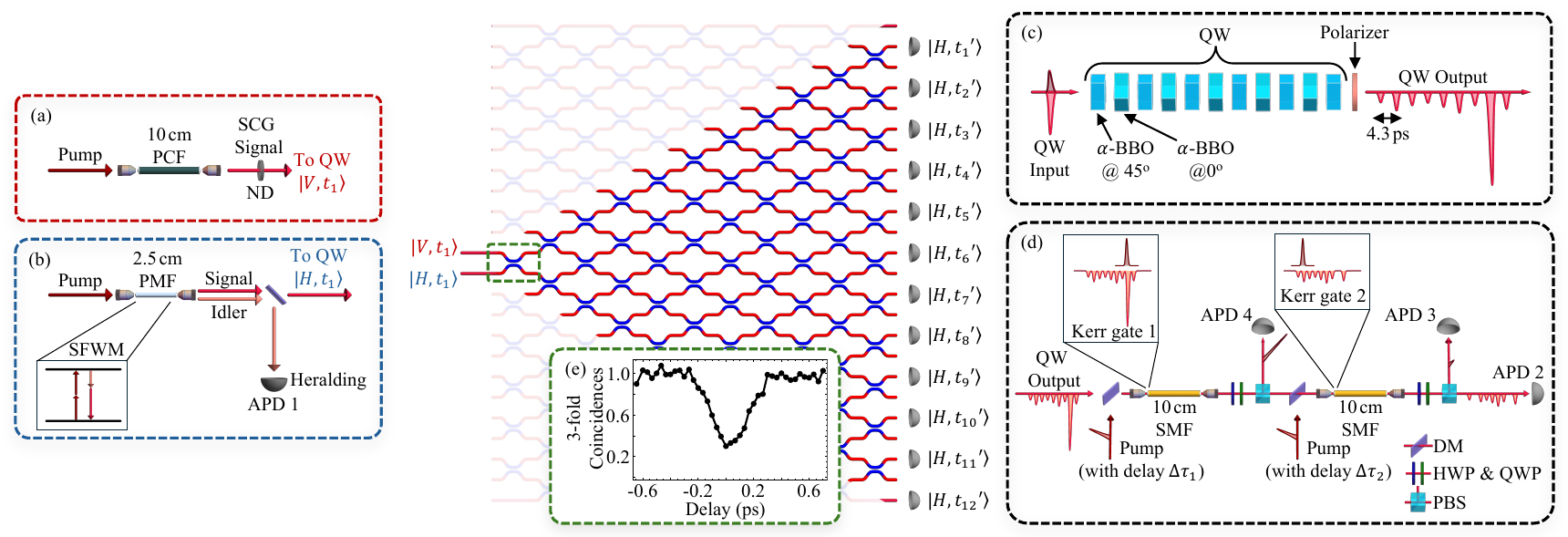}
	\caption{\textbf{Simplified experimental setup.} Photons from (a) an attenuated coherent state and (b) a heralded single-photon source enter (c) the QW in modes $\ket{V,t_1}$ and $\ket{H,t_1}$, respectively. The analogous planar circuit for (c) is shown in the centre background. Shown in inset (a), the attenuated coherent state is achieved through supercontinuum generation (SCG) in a 10\,cm photonic crystal fiber (PCF), which is spectrally filtered and attenuated to the single-photon level with neutral density (ND) filtering. Inset (b) depicts the heralded single-photon source, where photon pairs are generated through spontaneous four-wave mixing (SFWM) in 2.5\,cm of polarization-maintaining fiber (PMF)~\cite{smith2009photon,erskine2018real}. The idler photon ($\lambda_i=954$\,nm) is measured by the first avalanche photodiode (APD 1) and heralds the arrival of the signal photon ($\lambda_s=682$\,nm) at the output of the QW. Photons from these two sources traverse the $N$-layer QW ($N = 1 ... 11$), where each layer consists of a birefringent $\alpha$-BBO crystal, as seen in inset (c), with three rotational degrees of freedom. The multi-mode probability distribution across the possible horizontally polarized output states, $\ket{H,t_m\,'}$, is demultiplexed via optical Kerr gating, as shown in inset (d). In each optical Kerr gate, a strong pump pulse induces polarization rotation only where it overlaps in time with the QW output signal. Scanning the pump delay, $\Delta\tau_{1}$ and $\Delta\tau_{2}$, allows for readout of the entire horizontally polarized QW output state. Various coincidence patterns, measured by APDs 1–4 and processed using a time-to-digital converter, are analyzed to extract one-, two-, and three-photon interference. For example, detecting coincidences between APD 1, 2 and 4 (for a single interferometric layer, $N=1$, and when both Kerr gates are turned off) allows us to measure the Hong-Ou-Mandel (HOM) two-photon interference between the two different input states, as seen in inset (e), where coincidence counts are normalized. We observe up to 70\% HOM interference visibility between the two independent sources. DM: dichroic mirror, PBS: polarizing beam splitter, HWP: half-waveplate, QWP: quarter-waveplate, SMF: single mode fiber. See Fig. S1 in the Supplementary Materials for further experimental details.}
	\label{fig:concept}
\end{figure*}

One promising QW platform~\cite{fenwick2024photonic} which addresses these concerns uses ultrafast time-bin encoding (UTBE)~\cite{wang2024efficient,xu2018measuring,bouchard2023measuring,bouchard2024programmable}, where the beam splitter action occurs in the time domain, on ultrafast timescales. Sub-picosecond-duration photons enter and exit the time-domain beam splitter in time-bin modes, separated by several picoseconds. These time bins can be achieved, manipulated, and detected using birefringent crystals and ultrashort pulses~\cite{kupchak2017time,bouchard2022quantum,donohue2013coherent}; this enables operation in a single spatial mode, which offers high phase stability in transmission and compatibility with single-mode fibers. Direct measurement of the multi-mode time-bin signal at these time scales is not yet possible with fast photodetection and electronics; however, demultiplexing is possible with ultrafast optical Kerr gating~\cite{england2021perspectives}.

In this work, we implement a QW in a single spatial mode, using UTBE, to investigate one-, two-, and three-photon interference. Optical Kerr gating, with efficiency $>97\%$, is used to demultiplex the time-bin encoded probability distribution at the output of the QW. After demultiplexing, various click patterns at the detectors are recorded, allowing for the measurement of one-, two-, and three-photon probability distributions. Experimental measurements are compared to a simulation~\cite{QWCode} of the QW, which uses the open source quantum optics simulation package, Strawberry Fields~\cite{killoran2019strawberry,bromley2020applications}. Results demonstrate the versatility of the QW for generating, manipulating, and observing interference in a large interferometric network––a crucial capability in the context of photonic quantum information processing. Moreover, this experimental demonstration of multiphoton interference in a QW is consistent with requirements for quantum state generation~\cite{konno2024logical}, quantum state detection~\cite{pezze2007phase}, and quantum simulation~\cite{somhorst2023quantum}. 

\section*{Experimental Implementation}
The interferometric network, depicted in Fig.~\ref{fig:concept}, is based on a QW scheme consisting of layers of beam splitters. In general, a QW involves some quantum particle (or particles) with two degrees of freedom, represented by a tensor product of two Hilbert spaces. One Hilbert space is used to describe the walker, while the other is used to describe a coin. At each beam splitter layer, the particle(s) will have some probability of exiting in each of the possible output walker states, depending on how the coin portion of its state is prepared. Here, the quantum particles traversing the QW are photons from either an attenuated ultrafast laser pulse, depicted in Fig.~\ref{fig:concept}(a), or from a heralded single-photon source, depicted in Fig.~\ref{fig:concept}(b). Ultrafast time bins form the basis states of the walker, $\ket{t_m} \in \{ \ket{t_1} , \ket{t_2} , \ket{t_3} , ... \}$, while the polarization state of each photon defines its coin, with basis states $\ket{H}$ and $\ket{V}$, corresponding to horizontal and vertical polarization, respectively. The state of each photon is given by a tensor product of basis states from each of the walker and coin Hilbert spaces such that there are no initial correlations or entanglement between the degrees of freedom. 

Here, each beam splitter layer of the interferometric network corresponds to a single step of the photonic QW. The $n^{\text{th}}$ step in the QW is described by the unitary step operator, 
\begin{equation}
    \hat{U}_n = \hat{S}_n \cdot \hat{C}_n. 
    \label{eq:step}
\end{equation}
The shift operator, $\hat{S}_n$, raises or maintains the time-bin state of the photon, depending on its polarization state, and is given by
\begin{eqnarray}    \hat{S}_n=\sum_{m}\ket{H}\bra{H}\otimes\ket{t_m}\bra{t_m}+\ket{V}\bra{V}\otimes\ket{t_{m+1}}\bra{t_m}.
\end{eqnarray}
The coin operator is given by
\begin{eqnarray}
    \hat{C}_n=\sum_{m} \left(\text{cos}\frac{\Omega_n}{2}\ket{H}\bra{H}+\text{e}^{i \gamma_n}\text{sin}\frac{\Omega_n}{2}\ket{H}\bra{V}+ \nonumber\right. \\
    \left.\text{e}^{-i \gamma_n}\text{sin}\frac{\Omega_n}{2}\ket{V}\bra{H}-\text{cos}\frac{\Omega_n}{2}\ket{V}\bra{V}\right)\otimes\ket{t_m}\bra{t_m},
\end{eqnarray}
where the parameters, $\Omega_n$ and $\gamma_n$, can be adjusted to tune the probability amplitude and phase, respectively, of the $n^{\text{th}}$ step in the QW.

In this work, the unitary step operator in Eq.~\ref{eq:step} is achieved using a single birefringent $\alpha$-barium borate ($\alpha$-BBO) crystal. With the proper anti-reflection coating, the loss through each crystal is $< -0.045$\,dB. The temporal separation between ultrafast time bins is set by the thickness of the $\alpha$-BBO crystals; here, a crystal thickness of 10\,mm provides a temporal separation of $\sim$4.3\,ps at the signal wavelength. Each $\alpha$-BBO crystal can be rotated along three axes, providing individual programmability of each layer in the QW. Rotating the $n^\mathrm{th}$ $\alpha$-BBO crystal about the optical axis sets $\Omega_n$, while adjusting its tip and tilt sets $\gamma_n$.

Shown in Fig.~\ref{fig:concept}(c), the QW comprises a sequence of $N$ birefringent $\alpha$-BBO crystals, each performing one step of the walk, $\hat{U}_n$. The $N$-step QW can then be described by the unitary operator, 
\begin{equation}
    \hat{\mathscr{U}} = \prod^N_{n=1} \hat{U}_n,
\end{equation}
which acts on an initial probability amplitude vector. When more than one photon is input to the walk, it becomes easier to consider the QW evolution using the second quantization formalism. Here, we can count the number of photons in the $i^{\text{th}}$ mode, which is defined by a specific polarization (H or V) and time bin ($t_m$). In this picture, $\hat{\mathscr{U}}$ still provides all the information required to predict multiphoton evolution, but it now acts on each specific mode. For example, the case of a separable multiphoton input state with a total of $K$ photons can be written as
\begin{equation}
    \ket{\psi_{\text{in}}} = \prod^M_{i=1} \frac{1}{\sqrt{k_i !}} (\hat{a}^\dagger_i)^{k_i} \ket{0},
\end{equation}
where $\ket{0}$ is the  vacuum state, $(\hat{a}^\dagger_i)$ is the creation operator for a photon in mode $i$, and the total number of photons satisfies $K = \sum^M_{i=1} k_i$. The QW acts on the multimode creation operator such that it maps $\hat{a}^\dagger_i\rightarrow \hat{\mathscr{U}} \hat{a}^\dagger_i \hat{\mathscr{U}}^{\dagger}$. The total evolution of the $N$-step walk is then given by
\begin{equation}
    \ket{\psi^{(N)}_{\text{out}}} = \prod^M_{i=1} \frac{1}{\sqrt{k_i !}} (\hat{\mathscr{U}} \hat{a}^\dagger_i \hat{\mathscr{U}}^{\dagger})^{k_i} \ket{0}.
\end{equation}

Multiphoton interference will influence the measured probability distribution for a given output state, $\ket{\psi^{(N)}_{\text{out}}}$. Since the time bins defining the position of each walker occur on ultrafast timescales, ultrafast optical gating is necessary for output signal demultiplexing. QWs based on UTBE have been demonstrated with upconversion schemes for detection~\cite{xu2018measuring,wang2024efficient}; however, upconversion efficiency is often very low. Instead, we implement all-optical Kerr gating~\cite{kupchak2019terahertz,england2021perspectives}, depicted in Fig.~\ref{fig:concept}(d), for measurement. Kerr gating relies on a third-order $(\chi^{(3)})$ optical nonlinearity, by which a strong pulse of light is used to generate an intensity dependent refractive index modulation in the medium through which it propagates. This induces a birefringence, localized to the timing of the strong pulse, which can be used to rotate the polarization of a secondary optical signal, when the two are overlapped in the Kerr medium. Here, we use short (10\,cm) pieces of single-mode fiber (SMF) as the Kerr medium. Each Kerr gate is operated with a high efficiency of $>97\%$, which is crucial for observing multi-photon interference. 

An ideal measurement would simultaneously detect the photon number content of each time bin at the output of the multiphoton QW~\cite{goldberg2024correlations}; however, this is not possible with available technologies. Instead, we are able to extract information by gating out time bins individually (\textit{e.g.,} a one-fold measurement of time bin $t_{m_1}$ at APD 3) and pair-wise (\textit{e.g.,} a two-fold measurement of time bins $t_{m_1}$ and $t_{m_2}$ at APDs 3 and 4) with two Kerr gates. Time-resolved detection events in the gated time bins can also be measured in coincidence with detection events in the remaining time bins (\textit{e.g.,} a three-fold measurement of time bins $t_{m_1}$, $t_{m_2}$, and $[t_{m_3}]$, at APDs 3, 4, and 2, where $[t_{m_3}]$ refers to all remaining time bins, $[t_{m_3}]=\{t_m \text{ } | \text{ } m \neq {m_1} \land m\neq {m_2}\}$). We compare our measured detection probability patterns for heralded (conditional on an event at APD 1) and unheralded one, two, and three-fold events with those simulated using Strawberry Fields software.

\section*{Results}
\begin{figure}[t!]
	\centering
		\includegraphics[width=0.49\textwidth]{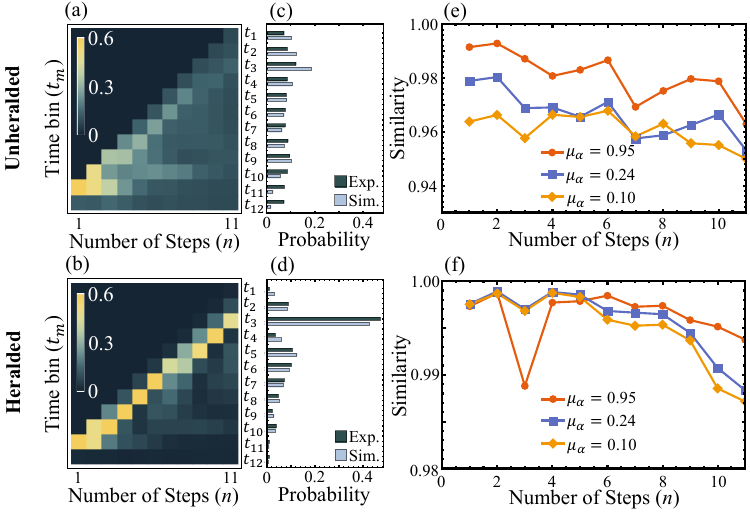}
	\caption{\textbf{One-fold measurements.} Unheralded and heralded one-fold measurements presented in the top and bottom rows, respectively. The step-by-step QW output evolution (for $\mu_{\alpha}=0.1$) is plotted in the first column, (a) and (b), with the corresponding 11-step probability distribution shown in the middle column, (c) and (d). Experimental data is compared with a simulation through a similarity calculation, plotted in the last column, (e) and (f), for the three different values of $\mu_{\alpha}$.}
	\label{fig:onePhoton}
\end{figure}

In order to observe the effect of multiphoton quantum interference in the QW, we use two distinct categories of input state to the network. The signal mode from the two-mode squeezed vacuum (TMSV) source (Fig.~\ref{fig:concept}(b)) is input to the $\ket{H,t_1}$ bin for all measurements. This acts as a nonclassical source of heralded single photons when APD 1 detects an idler photon. In the unheralded case, when all laser shots are included, the input is a single-mode thermal state, with mean photon number $\mu_\xi\approx0.026$. Either of these two input states is input to the QW along with an attenuated coherent state (ACS), with mean photon number $\mu_\alpha$, always input to $\ket{V,t_1}$. We measure the output QW probability distribution across the possible horizontally polarized output time bins and examine one-, two-, and three-fold coincidences for both the heralded (single photon input) and unheralded (thermal state input) cases. See Table S1 in the Supplementary Materials for details on the various coincidence patterns considered. Data is collected for three different values of $\mu_{\alpha}$ in order to observe its effects. 

Quantum interference requires that the photons in the walk are indistinguishable (\textit{i.e.,} ideally, with perfect mode overlap). We quantify this by measuring Hong-Ou-Mandel (HOM) interference between the ACS and the heralded single photon, for a single QW layer. For a given squeezing parameter of the heralded single-photon source, the HOM visibility will also depend on the mean photon number of the ACS~\cite{rarity2005non,xu2023characterization}. We measure a maximum HOM visibility of 70\%, which is used as an estimate for the mode overlap between the two input sources in the Strawberry Fields simulations.

\begin{figure*}[t!]
	\centering
		\includegraphics[width=1\textwidth]{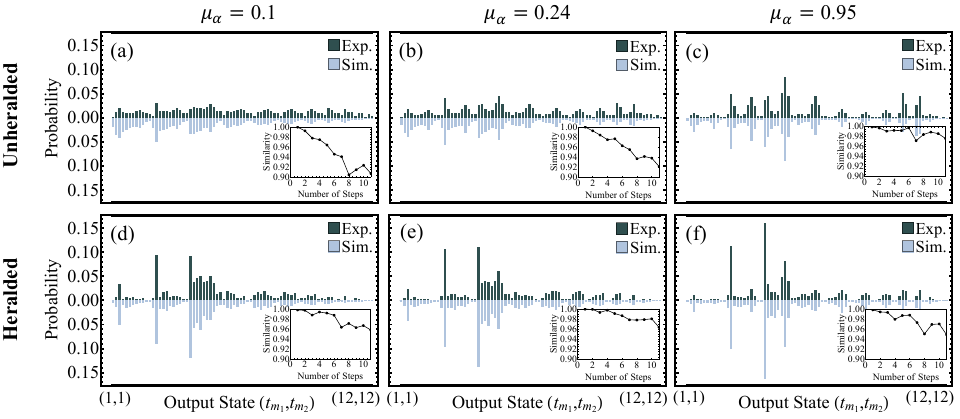}
	\caption{\textbf{Two-fold measurements.} Unheralded and heralded measurements of the QW output distributions from the 11-step QW are shown in the top, (a)-(c), and bottom, (d)-(f), rows, respectively. Data is collected for $\mu_{\alpha}=$0.1 (first column), 0.24 (second column), and 0.95 (third column). The step-by-step similarities between each experimental dataset and its corresponding simulation are shown in the respective insets.}
	\label{fig:twoPhoton}
\end{figure*}

We first consider the heralded and unheralded one-fold measurements of the QW output, as shown in Fig.~\ref{fig:onePhoton}. Here, only one Kerr gate and APD 4 is required to make the measurement. The unheralded and heralded distributions are shown in Figs.~\ref{fig:onePhoton}(a) and (b), respectively. The QW output after 11 steps is plotted in Figs.~\ref{fig:onePhoton}(c) and (d), next to the corresponding step-by-step QW evolution from (a) and (b), respectively. Here, experimental data is plotted next to the Strawberry Fields simulation~\cite{QWCode}, for the corresponding photonic circuit. Each step-by-step QW output probability distribution is measured for three values of $\mu_{\alpha}$, and a similarity calculation between experiment and simulation is made using the Bhattacharyya coefficient. Similarity calculations are presented in Figs.~\ref{fig:onePhoton}(c) and (d), and correspond to the data presented in (a) and (b), respectively. In all data presented, the similarity remains above 95\%. The qualitative difference between the heralded and unheralded distributions is determined by the relative brightness of the light at the two input ports~\cite{fenwick2024photonic}. For the unheralded case, the output distribution is dominated by the ACS $V$-polarized input ($\mu_\alpha\gg\mu_\xi$); for the heralded case, the output distribution is dominated by the $H$-polarized single photon input ($\mu_\alpha\ll1$).

Next, heralded and unheralded two-fold measurements are presented in Fig.~\ref{fig:twoPhoton}. Here, two Kerr gates and APDs 3 and 4 are required to measure the complete probability distribution. The unheralded and heralded two-fold measurements are presented in the top and bottom rows, respectively. Measurements are taken for $\mu_{\alpha}=$ 0.1, 0.24, and 0.95. Each experimental dataset is compared with the simulation, through a similarity calculation, as shown in the respective insets in Fig.~\ref{fig:twoPhoton}. This simulation propagates an ACS and the signal mode of a TMSV state through $N$ steps of the QW unitary in Eq.~\ref{eq:step}. Details on the simulation are provided in the Supplementary Materials (see Fig. S5). 

We see that the probability distributions in Figs.~\ref{fig:twoPhoton}(d)-(f), for which the two-fold measurements are heralded, show stronger clustering than the distributions in Figs.~\ref{fig:twoPhoton}(a)-(c), for which the two-fold measurements are not heralded. For example, the largest probabilities measured in (d), (e), and (f) are 3.0, 2.4, and 1.9 times larger than the largest probabilities measured in (a), (b), and (c), respectively. The probability distributions in (d)–(f) also show detection outcomes which are more suppressed overall. Both of these effects are signatures of bosonic quantum interference, caused by nonclassical input states. Another way to investigate this nonclassicality is by testing the accuracy of an optimized classical model. This model simulates the experiment using a two-mode squeezed thermal (TMST) source, also referred to as a ``squashed state'', rather than the TMSV source. For low enough squeezing, the TMST source provides the closest classical approximation to a photon pair source after loss~\cite{qi2020regimes}. We find significantly better similarities when experimental data is compared with the quantum simulation using a TMSV source, rather than a TMST source. See Fig. S6 in the Supplementary Materials for details on this comparison. 

\begin{figure}[t!]
	\centering
		\includegraphics[width=0.49\textwidth]{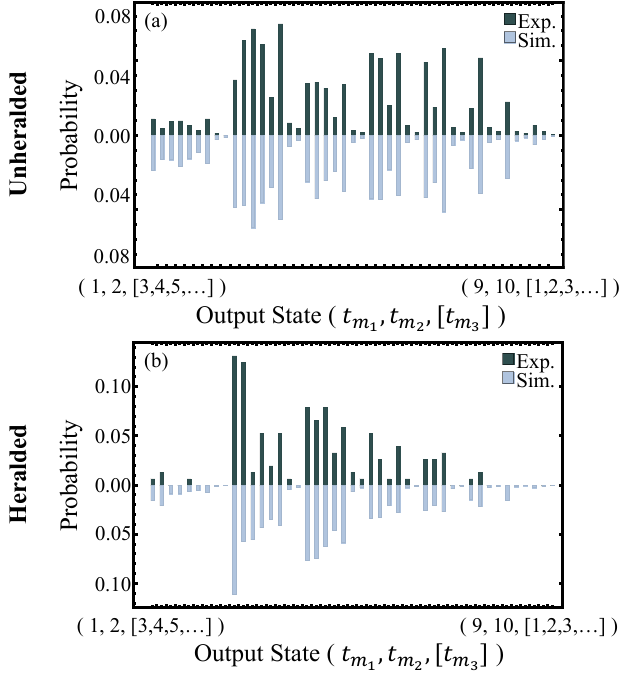}
	\caption{\textbf{Partial three-fold measurements.} Output distributions from a 9-step QW are shown when three-fold measurements are (a) unheralded and (b) heralded. Data is shown for $\mu_{\alpha}=0.95$. Experimental data is compared to simulation, with similarities of (a) 95\% and (b) 89\%, respectively.}
	\label{fig:threePhoton}
\end{figure}

Finally, we present heralded and unheralded three-fold measurements in Fig.~\ref{fig:threePhoton}. Due to the fact that only two Kerr gates are implemented in this work, we are limited to measurements where one detector measures an event in time bin $t_{m_1}$, another detector measures an event in time bin $t_{m_2}$, and the remaining detector measures an event in time bin $[t_{m_3}]$, where $[t_{m_3}]=\{t_m \text{ } | \text{ } m \neq {m_1} \land m\neq {m_2}\}$. This approach is possible only because the Kerr gates are operated with such high efficiency, ensuring that photons in time bins $t_{m_1}$ and $t_{m_2}$ have a very low probability of arriving at the detector which measures $[t_{m_3}]$. See Fig. S7 for a depiction of this experimental configuration. We note here that with a third Kerr gate, the full three-fold probability distribution could be measured directly. Nevertheless, we present a subset of the unheralded and heralded three-fold measurements at the output of a 9-step QW in Fig.~\ref{fig:threePhoton}(a) and (b), respectively. The similarities between experiment and simulation are 95\% and 89\% for the unheralded and heralded three-fold measurements, respectively. We note that all measurements in Fig.~\ref{fig:threePhoton} correspond to a value of $\mu_{\alpha}=0.95$, which provides the highest coincidence rates. A similarity calculation for the step-by-step evolution of the three-fold measurements is provided in Fig. S8.

\section*{Discussion and outlook}
The QW model, which is essentially a large, multi-mode interferometric network, provides a convenient testbed for interference-based quantum platforms (\textit{e.g.,} Gaussian boson sampling~\cite{thekkadath2022experimental}), for example in quantum information processing and quantum metrology. By implementing a QW here, we are able to demonstrate the potential of UTBE for the generation, manipulation, and measurement of multi-mode interference between up to three photons in a single spatial mode. This is a key step towards scalable quantum information processing platforms, which require high-dimensional multi-mode interference of many photons in order to push beyond the current limitations of classical computation, entering the realm of ``useful'' quantum computation. 

In practice, many-photon operation still poses a large challenge. One of the main limitations for multi-photon operation in many-mode interferometric networks is that the loss scales with the circuit depth. As such, achieving low loss per circuit depth, as demonstrated here, is crucial to scale these photonic circuits to the multi-photon regime. Single-photon sources pose another limitation to the scalability of photonic quantum information processing schemes. Deterministic sources, such as quantum dots~\cite{laferriere2022unity} or defect centers~\cite{mizuochi2012electrically} are promising for on-demand single photons with high rates; however, achieving high interference visibility is still difficult. Probabilistic sources, generated through nonlinear processes such as the SFWM source in this work or spontaneous parametric downconversion~\cite{hong1987measurement} can be used to achieve high interference visibility, but due to their probabilistic nature pose constraints on count rates; however, with photon number resolving detectors, probabilistic sources can be operated at higher rates, making them a promising avenue for photonic quantum information processing~\cite{deng2023gaussian}. This would be compatible with continuous-variable (CV) quantum information processing schemes, which consist of deterministic sources of squeezed light~\cite{andersen2010continuous,yonezawa2010continuous}. 

Not only useful for information processing, quantum interference holds implications for metrology where the large interferometric network presented here could be extended for parameter estimation~\cite{ellinas2024parameter,rajauria2020estimation,annabestani2022multiparameter}. Most proposals of parameter estimation with QWs consider the case where the parameters of the walk itself are being estimated. This is useful in the context of implementing QWs with low experimental error; parameter estimation could be used to determine which, if any, of the many QW parameters require adjustment. Parameter estimation in a QW may also be extended to other applications including, for example, the analysis of gravitational wave measurements~\cite{escrig2023parameter}.

With the demonstrated ability to handle interference of up to three photons across 24 modes, our QW with UTBE holds potential in interference-based quantum platforms. We anticipate that the QW will be a useful tool in quantum metrology, and, with improvement to multiphoton event detection rates, will be capable of manipulating interference between $>3$ photons, providing a pathway for scalable photonic quantum information processing.

\section*{Acknowledgments}
This work is supported by the Vanier Canada Graduate Scholarships program. We thank Alicia Sit, Andrew Proppe, Brayden Freitas, Denis Guay, Doug Moffatt, Milica Banic, Nicolas Couture, Noah Lupu-Gladstein, Ramy Tannous, Rune Lausten, and Yingwen Zhang for their support and insightful discussions.


\providecommand{\noopsort}[1]{}

\end{document}